# Enhancement of $T_c$ by uniaxial lattice contraction in BiS$_2$-based superconductor PrO$_{0.5}$F$_{0.5}$BiS$_2$

Joe Kajitani[1], Keita Deguchi[2], Takafumi Hiroi[1], Atsushi Omachi[1], Satoshi Demura[2], Yoshihiko Takano[2], Osuke Miura[1], and Yoshikazu Mizuguchi[1, 2]*

Recently, several types of BiS$_2$-based superconductor such as Bi$_4$O$_4$S$_3$,[1] REO$_{1-x}$F$_x$BiS$_2$ (RE: rare earth),[2-6] and Sr$_{1-x}$La$_x$FBiS$_2$ [7,8] have been discovered. The typical parent material LaOBiS$_2$ has a crystal structure composed of an alternate stacking of BiS$_2$ conduction layers and LaO blocking layers. Electron carriers, which are essential for the appearance of superconductivity in the BiS$_2$-based family, can be controlled by manipulating the structure and composition at the blocking layers. In LaOBiS$_2$, electron carriers can be generated by a partial substitution of O$^{2-}$ with F$^-$.[2,9]

The superconducting properties of the BiS$_2$-based superconductors strongly depend on the sample preparation method. Solid-state-reacted LaO$_{1-x}$F$_x$BiS$_2$ shows filamentary superconductivity; in other words, its superconducting volume fraction is low, whereas its superconducting states are evidently generated. To induce bulk superconductivity, high-pressure annealing [2,10] or application of high-pressure [11-13] is effective. Using the high-pressure technique, the transition temperature ($T_c$) of LaO$_{0.5}$F$_{0.5}$BiS$_2$ reaches 10.6 K. It was found that the crystal structure obviously changes after high-pressure annealing. Recently, we have reported that the uniaxial lattice contraction generated along the $c$ axis could be positively linked to the enhancement of superconductivity in high-pressure-annealed LaO$_{1-x}$F$_x$BiS$_2$.[14]

In this study, we focus on PrO$_{0.5}$F$_{0.5}$BiS$_2$. It was reported that solid-state-reacted PrO$_{0.5}$F$_{0.5}$BiS$_2$ shows bulk superconductivity with a $T_c$ of 3.5 K.[5] Furthermore, its $T_c$ determined by the electrical resistivity measurement under high pressure reaches a maximum of 7.6 K.[11] However, the correlation between the superconducting properties and the crystal structure under high pressure in $PrO_{0.5}F_{0.5}BiS_2$ remains to be clarified. Here, we show that the $T_c$ of $PrO_{0.5}F_{0.5}BiS_2$ is increased by high-pressure annealing, and that uniaxial contraction along the $c$ axis observed after high-pressure annealing correlates with the increase in $T_c$.

Polycrystalline samples of $PrO_{0.5}F_{0.5}BiS_2$ were prepared by a solid-state reaction using powders of $Bi_2O_3$ (99.9 %), $BiF_3$ (99.9 %), $Bi_2S_3$, $Pr_2S_3$, and grains of Bi (99.99 %). The $Bi_2S_3$ powder was prepared by reacting Bi (99.99 %) and S (99.99 %) grains in an evacuated quartz tube. The $Pr_2S_3$ powder was prepared by reacting Pr (99.9 %) and S (99.99%) grains in an evacuated quartz tube. Other chemicals used were purchased from Kojundo-Kagaku Laboratory. The starting materials with a nominal composition of $PrO_{0.5}F_{0.5}BiS_2$ were well-mixed, pressed into pellets, sealed in an evacuated quartz tube, and heated at 700 ºC for 10 h; in this article, we call this sample the "As-grown" sample. The obtained products were ground, sealed in an evacuated quartz tube and heated again under the same heating conditions to obtain a homogenized sample. The obtained sample was annealed at 600 ºC under high pressure of 3 G Pa for 1 h using a cubic-anvil high-pressure synthesis instrument; in this article, we call this material the "HP" sample. All the obtained samples were characterized by X-ray diffraction analysis using the $\theta$–$2\theta$ method with CuK$\alpha$ radiation. The temperature dependence of electrical resistivity was measured using the four-terminal method. The temperature dependence of magnetic susceptibility was measured using a superconducting quantum interface device (SQUID) magnetometer with an applied field of 5 Oe after both zero-field cooling (ZFC) and field cooling (FC).

Figure 1(a) shows the X-ray diffraction patterns of the As-grown and HP samples. Almost all of the obtained peaks were explained using the tetragonal *P4/nmm* space group. The numbers displayed in the profile indicate Miller indices. The obtained lattice constants were $a$ = 4.0323 Å and $c$ = 13.5038 Å for the As-grown sample, and $a$ = 4.0323 Å and $c$ = 13.4587 Å for the HP sample. The obtained X-ray diffraction patterns are almost the same, but the peaks of the HP sample slightly shift and are broadened. To discuss the change in the lattice constants, the enlarged X-ray profiles around the (004) and (200) peaks are shown in Fig. 1(b) and 1(c), respectively. The (004) peak of the HP sample shifts to higher angles, which clearly indicates the contraction of the *c* axis with high-pressure annealing. In contrast, the (200) peak position does not change. Namely, the *a* axis does not show a remarkable change with high-pressure annealing. Nevertheless, it can note that the (200) peak of the HP sample is slightly broadened toward lower angles, which indicates that the *a* axis is slightly strained(expanded) with high-pressure annealing. These suggest that the *c* axis is easily compressed, while the *a* axis is stable under high pressure. Therefore, the uniaxial contraction along the *c* axis is induced after high-pressure annealing in PrO$_{0.5}$F$_{0.5}$BiS$_2$, as observed in the LaO$_{0.5}$F$_{0.5}$BiS$_2$ system.

Figure 2(a) shows the temperature dependences of electrical resistivity for the As-grown and HP samples. For the As-grown sample, the resistivity decreases with decreasing temperature between 250 and 300 K, then slightly increases with decreasing temperature below 250 K. For the HP sample, the resistivity increases with decreasing temperature. Therefore, semiconducting characteristics are relatively enhanced by high-pressure annealing. Figure 2(b) shows the temperature dependences of electrical resistivity for the As-grown and HP samples below 15 K. The As-grown sample shows transition temperatures of $T_c^{onset}$ = 4.0 K and $T_c^{zero}$ = 3.6 K. The HP sample shows transition temperatures of $T_c^{onset}$ = 9.4 K and $T_c^{zero}$ = 5.5 K. $T_c^{onset}$ was defined as the temperature where the resistivity deviated from the extrapolation line, as indicated by arrows.

Figure 3 shows the temperature dependences of magnetic susceptibility (ZFC) for the As-grown and HP samples. The As-grown sample shows a $T_c^{mag}$ of 3.5 K, while the HP sample shows a $T_c^{mag}$ of 5.0 K. A large shielding volume fraction, $\chi < -1/4\pi$, is observed in the ZFC data at 2 K for both samples, indicating that both samples show bulk superconductivity.

Here, we briefly discuss why $T_c$ is increased in the HP sample. In the LaO$_{0.5}$F$_{0.5}$BiS$_2$ system, it was suggested that the increase in $T_c$ is observed when the crystal structure changed from tetragonal to monoclinic under high pressure.[13] If the crystal structure of the HP sample is monoclinic, we should observe the split of both the (200) and (004) peaks of the tetragonal As-grown sample. As shown in Fig. 1, these peaks of the HP sample are slightly broadened, but they do not split. Therefore, we assume that the crystal structure of the HP-annealed PrO$_{0.5}$F$_{0.5}$BiS$_2$ with a higher $T_c$ is basically tetragonal, and that a uniaxial lattice contraction along the $c$ axis is essential for the increase in $T_c$. It was predicted that the $z$ coordinate of the in-plane S site could easily change with a change in F concentration and that the band structure significantly changes because of the change of the $z$ coordinate of the S site.[15,16] Recently, crystal structure analysis using $LaO_{1-x}F_xBiS_2$ single crystals has revealed that the distortion of the Bi-S plane disappears with increasing F concentration.[17] On the basis of these previous reports, we assume that the atomic position of the in-plane S site can be optimized by HP annealing, which results in an increase in $T_c$ in HP-annealed $PrO_{0.5}F_{0.5}BiS_2$.

In summary, we investigated the crystal structure and superconducting properties of As-grown and high-pressure-annealed $PrO_{0.5}F_{0.5}BiS_2$. We found that high-pressure annealing generates uniaxial lattice contraction along the $c$ axis. Both As-grown and high-pressure-annealed $PrO_{0.5}F_{0.5}BiS_2$ show bulk superconductivity. The $T_c$ of $PrO_{0.5}F_{0.5}BiS_2$ is clearly increased from $T_c^{zero} = 3.6$ K to $T_c^{zero} = 5.5$ K by high-pressure annealing. Unexpectedly, the semiconducting characteristics are relatively enhanced by high-pressure annealing. Namely, we assume that the increase in $T_c$ cannot be explained by an increase in the number electron carriers. Having considered these facts, we conclude that the increase in $T_c$ correlates with the uniaxial lattice contraction along the $c$ axis in $PrO_{0.5}F_{0.5}BiS_2$.

**Acknowledgements**

This work was partly supported by a Grant-in-Aid for Scientific Research for Young Scientists (A).

*E-mail: mizugu@tmu.ac.jp

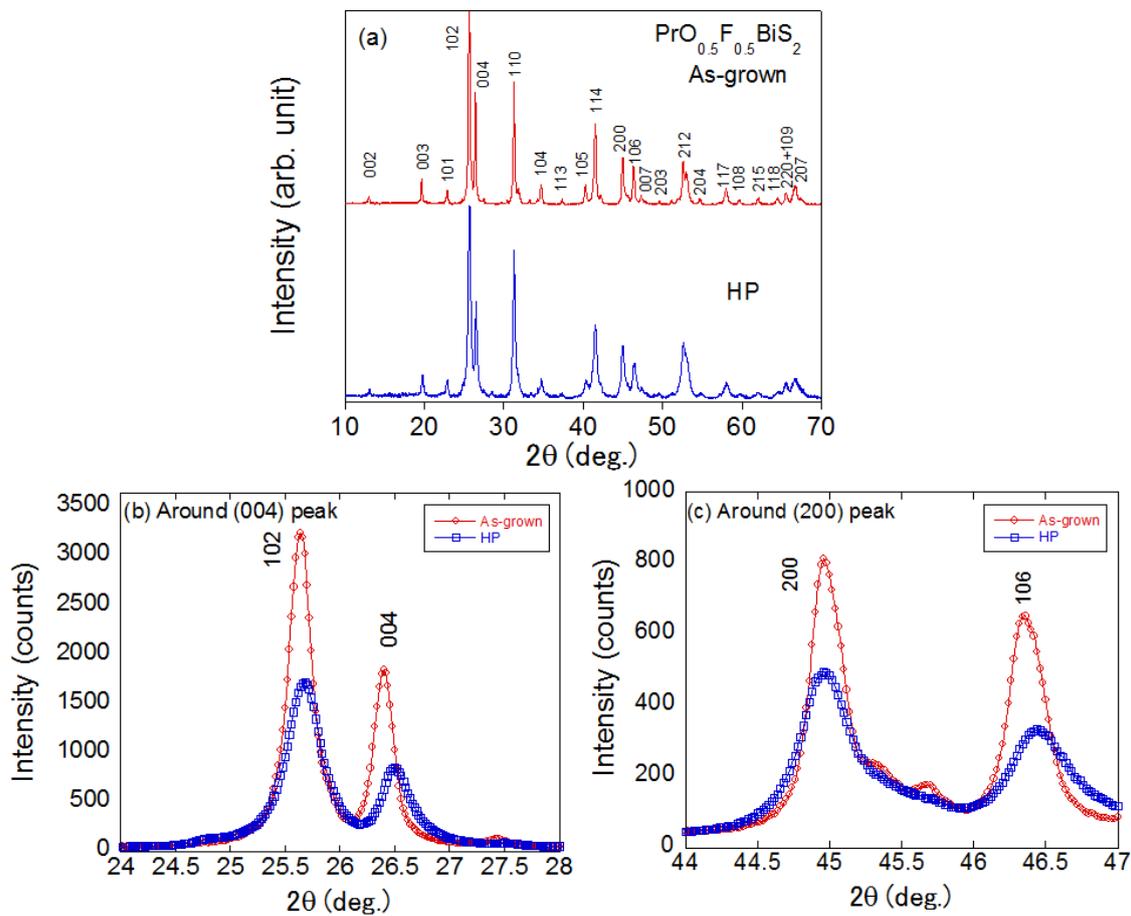

Fig. 1. (a) X-ray diffraction patterns of As-grown and HP samples. (b) Enlarged X-ray profiles near the (004) peaks of As-grown and HP samples. (c) Enlarged X-ray profiles near the (200) peaks of As-grown and HP samples.

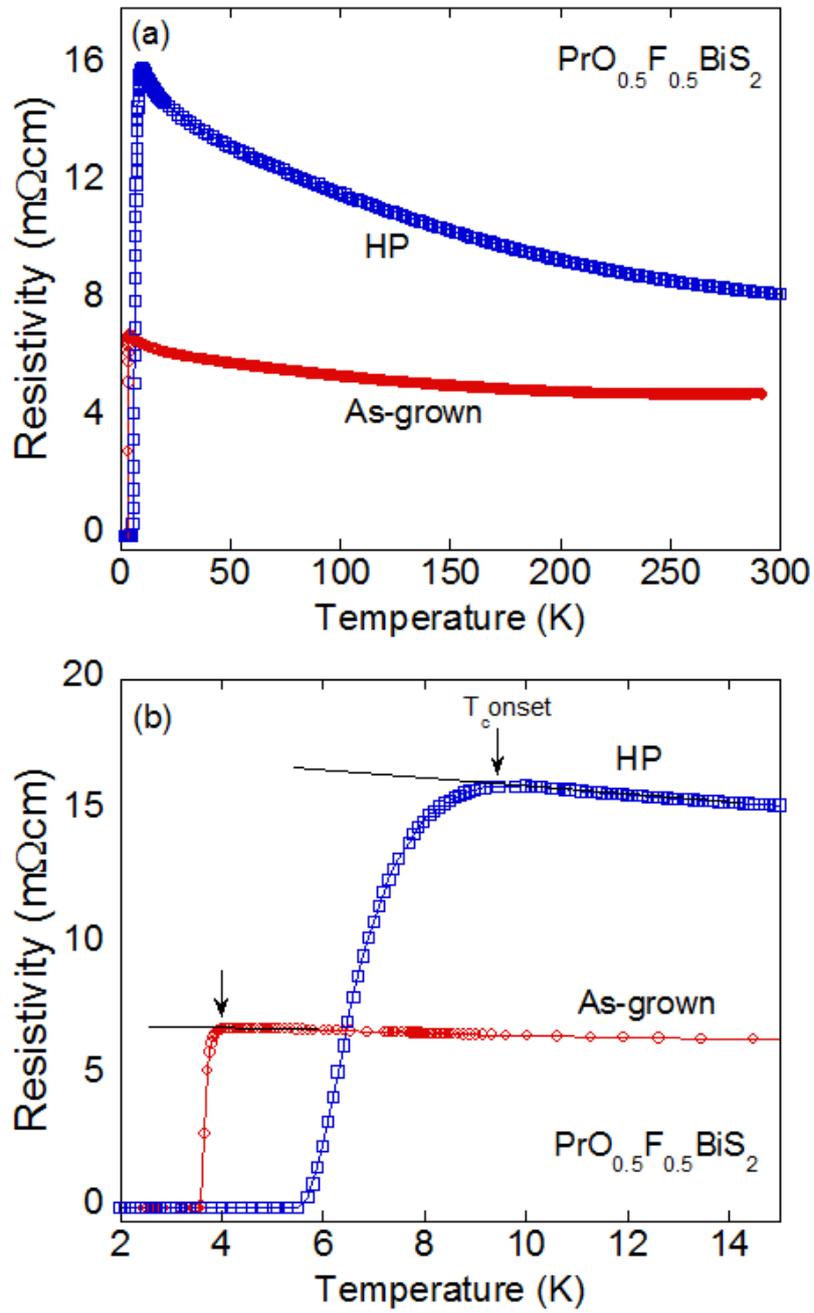

Fig. 2. (a) Temperature dependences of resistivity for As-grown and HP samples. (b) Temperature dependences of resistivity for As-grown and HP samples below 15 K.

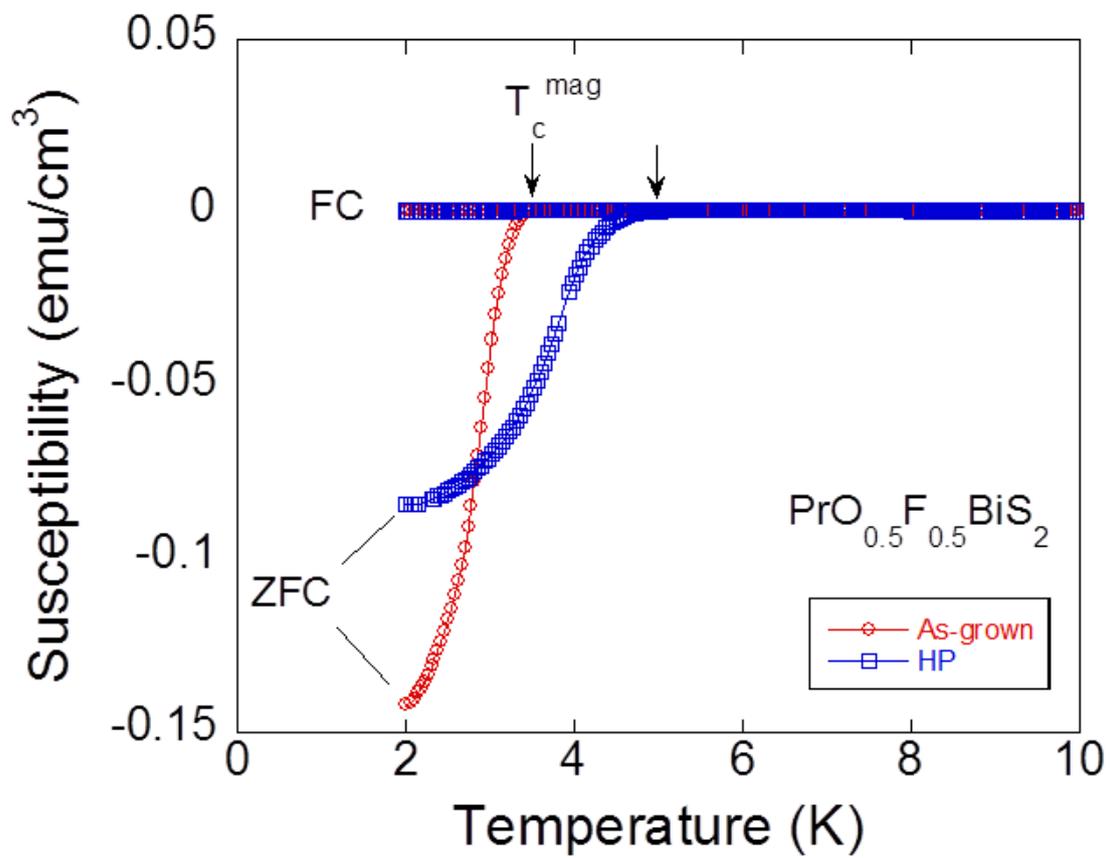

Fig. 3. Temperature dependences of magnetic susceptibility for As-grown and HP samples.